\documentclass[aps, prr, a4paper, reprint, superscriptaddress, floatfix]{revtex4-2}

\usepackage{amsmath, amssymb}
\usepackage{color, hyperref}
\usepackage[all]{hypcap} 
\usepackage{wrapfig}
\usepackage{graphicx}
\usepackage{xcolor}
\usepackage{placeins}
\usepackage[normalem]{ulem}
\usepackage{bm}

\definecolor{MyDarkGreen}{rgb}{0,0.6,0}
\definecolor{MyDarkBlue}{rgb}{0,0,0.8}
\definecolor{MyDarkRed}{rgb}{0.6,0,0.3}
\hypersetup{breaklinks=true,colorlinks=true,plainpages=true,linktocpage=true,linkcolor=MyDarkBlue,citecolor=MyDarkGreen,urlcolor=MyDarkRed,pdfborder={0 0 0}}

\newlength{\figurewidth}
\begin{document}

\title{Wavelength-Dependent Photodissociation of Iodomethylbutane}

% Authors, for the paper (add full first names)
\author{Valerija Music}
\affiliation{Department of Physics, Universit\"at Hamburg, 22607 Hamburg, Germany}
\affiliation{Institut f{\"u}r Physik und CINSaT, Universit{\"a}t Kassel, 34132 Kassel, Germany}
\affiliation{European X-Ray Free-Electron Laser Facility, 22869 Schenefeld, Germany}

\author{Felix Allum}
\affiliation{Stanford PULSE Institute, SLAC National Accelerator Laboratory, Menlo Park, California 94305, USA}
\affiliation{The Chemistry Research Laboratory, Department of Chemistry, University of Oxford, Oxford OX1 3TA, United Kingdom}

\author{Ludger Inhester} 
\affiliation{Center for Free-Electron Laser Science CFEL, Deutsches Elektronen-Synchrotron DESY, Notkestr. 85, 22607 Hamburg, Germany}

\author{Philipp Schmidt}
\affiliation{European X-Ray Free-Electron Laser Facility, 22869 Schenefeld, Germany}
\affiliation{Institut f{\"u}r Physik und CINSaT, Universit{\"a}t Kassel, 34132 Kassel, Germany}

\author{Rebecca Boll}
\affiliation{European X-Ray Free-Electron Laser Facility, 22869 Schenefeld, Germany}

\author{Thomas M. Baumann}
\affiliation{European X-Ray Free-Electron Laser Facility, 22869 Schenefeld, Germany}

\author{Günter Brenner}
\affiliation{Deutsches Elektronen-Synchrotron DESY, Notkestr. 85, 22607 Hamburg, Germany}

\author{Mark Brouard}
\affiliation{The Chemistry Research Laboratory, Department of Chemistry, University of Oxford, Oxford OX1 3TA, United Kingdom}

\author{Michael Burt}
\affiliation{The Chemistry Research Laboratory, Department of Chemistry, University of Oxford, Oxford OX1 3TA, United Kingdom}

\author{Philipp V. Demekhin}
\affiliation{Institut f{\"u}r Physik und CINSaT, Universit{\"a}t Kassel, 34132 Kassel, Germany}

\author{Simon D\"orner}
\affiliation{Deutsches Elektronen-Synchrotron DESY, Notkestr. 85, 22607 Hamburg, Germany}

\author{Arno Ehresmann}
\affiliation{Institut f{\"u}r Physik und CINSaT, Universit{\"a}t Kassel, 34132 Kassel, Germany}

\author{Andreas Galler}
\affiliation{European X-Ray Free-Electron Laser Facility, 22869 Schenefeld, Germany}

\author{Patrik Grychtol}
\affiliation{European X-Ray Free-Electron Laser Facility, 22869 Schenefeld, Germany}

\author{David Heathcote}
\affiliation{The Chemistry Research Laboratory, Department of Chemistry, University of Oxford, Oxford OX1 3TA, United Kingdom}

\author{Denis Kargin}
\affiliation{Institut für Chemie, Universität Kassel, Heinrich-Plett-Straße 40, 34132 Kassel, Germany}

\author{Mats Larsson}
\affiliation{Stockholm University, AlbaNova University Center, 114 21 Stockholm, Sweden}

\author{Jason W. L. Lee}
\affiliation{Deutsches Elektronen-Synchrotron DESY, Notkestr. 85, 22607 Hamburg, Germany}
\affiliation{The Chemistry Research Laboratory, Department of Chemistry, University of Oxford, Oxford OX1 3TA, United Kingdom}

\author{Zheng Li}
\affiliation{State Key Laboratory for Mesoscopic Physics, School of Physics, Peking University, Beijing 100871, China}
\affiliation{Collaborative Innovation Center of Extreme Optics, Shanxi University, Taiyuan, Shanxi 030006, China}

\author{Bastian Manschwetus}
\affiliation{Deutsches Elektronen-Synchrotron DESY, Notkestr. 85, 22607 Hamburg, Germany}

\author{Lutz Marder}
\affiliation{Institut f{\"u}r Physik und CINSaT, Universit{\"a}t Kassel, 34132 Kassel, Germany}

\author{Robert Mason}
\affiliation{The Chemistry Research Laboratory, Department of Chemistry, University of Oxford, Oxford OX1 3TA, United Kingdom}

\author{Michael Meyer}
\affiliation{European X-Ray Free-Electron Laser Facility, 22869 Schenefeld, Germany}

\author{Huda Otto}
\affiliation{Institut f{\"u}r Physik und CINSaT, Universit{\"a}t Kassel, 34132 Kassel, Germany}

\author{Christopher Passow}
\affiliation{Deutsches Elektronen-Synchrotron DESY, Notkestr. 85, 22607 Hamburg, Germany}

\author{Rudolf Pietschnig}
\affiliation{Institut für Chemie, Universität Kassel, Heinrich-Plett-Straße 40, 34132 Kassel, Germany}

\author{Daniel Ramm}
\affiliation{Deutsches Elektronen-Synchrotron DESY, Notkestr. 85, 22607 Hamburg, Germany}

\author{Daniel Rolles}
\affiliation{Kansas State University, 1228 N 17th St, KS 66506, United States of America}

\author{Kaja Schubert}
\affiliation{Deutsches Elektronen-Synchrotron DESY, Notkestr. 85, 22607 Hamburg, Germany}

\author{Lucas Schwob}
\affiliation{Deutsches Elektronen-Synchrotron DESY, Notkestr. 85, 22607 Hamburg, Germany}

\author{Richard D. Thomas}
\affiliation{Stockholm University, AlbaNova University Center, 114 21 Stockholm, Sweden}

\author{Claire Vallance}
\affiliation{The Chemistry Research Laboratory, Department of Chemistry, University of Oxford, Oxford OX1 3TA, United Kingdom}

\author{Igor Vidanovic}
\affiliation{Institut f{\"u}r Physik und CINSaT, Universit{\"a}t Kassel, 34132 Kassel, Germany}

\author{Clemens von Korff Schmising}
\affiliation{Max Born Institute, Max-Born-Straße 2A, 12489 Berlin, Germany}

\author{René Wagner}
\affiliation{European X-Ray Free-Electron Laser Facility, 22869 Schenefeld, Germany}

\author{Vitali Zhaunerchyk}
\affiliation{University of Gothenburg, 405 30 Gothenburg, Sweden}

\author{Peter Walter}
\affiliation{Department of Physics, Universit\"at Hamburg, 22607 Hamburg, Germany}
\affiliation{SLAC National Accelerator Laboratory, 2575 Sand Hill Road, Menlo Park, California 94025, USA}
\affiliation{TAU Systems, Austin, TX, USA}

\author{Sadia Bari}
\affiliation{Deutsches Elektronen-Synchrotron DESY, Notkestr. 85, 22607 Hamburg, Germany}

\author{Benjamin Erk}
\affiliation{Deutsches Elektronen-Synchrotron DESY, Notkestr. 85, 22607 Hamburg, Germany}

\author{Markus Ilchen} \email{markus.ilchen@uni-hamburg.de}
\affiliation{Department of Physics, Universit\"at Hamburg, 22607 Hamburg, Germany}
\affiliation{Center for Free-Electron Laser Science CFEL, Deutsches Elektronen-Synchrotron DESY, Notkestr. 85, 22607 Hamburg, Germany}
\affiliation{Deutsches Elektronen-Synchrotron DESY, Notkestr. 85, 22607 Hamburg, Germany}
\affiliation{European X-Ray Free-Electron Laser Facility, 22869 Schenefeld, Germany}
\affiliation{Institut f{\"u}r Physik und CINSaT, Universit{\"a}t Kassel, 34132 Kassel, Germany}

\begin{abstract}
Ultrashort XUV pulses of the Free-Electron-LASer in Hamburg (FLASH) were used to investigate laser-induced fragmentation patterns of the prototypical chiral molecule 1-iodo-2-methyl-butane (C$_5$H$_{11}$I) in a pump-probe scheme. Ion velocity-map images and mass spectra of optical-laser-induced fragmentation were obtained for subsequent FEL exposure with photon energies of 63\thinspace eV and 75\thinspace eV. These energies specifically address the iodine 4d edge of neutral and singly charged iodine, respectively.
The presented ion spectra for two optical pump-laser wavelengths, i.e., 800 nm and 267 nm, reveal substantially different cationic fragment yields in dependence on the wavelength and intensity. For the case of 800-nm-initiated fragmentation, the molecule dissociates notably slower than for the 267\thinspace nm pump. The results underscore the importance of considering optical-laser wavelength and intensity in the dissociation dynamics of this prototypical chiral molecule that is a promising candidate for future studies of its asymmetric nature.
\end{abstract}

\maketitle

\section{Introduction}

Photodissociation is an important restructuring process of matter with a broad interest in the biological, chemical, and physical sciences. The time-resolved investigation of photodissociation with element- or 'site'-specificity provides unique insight into the different dynamics and pathways of e.g., molecular fragmentation, the underlying processes of radiation damage, chemical bond-breaking, and light-matter interaction in general \cite{young2018roadmap}. These processes evolve to a significant extent on femtosecond timescales and are therefore only directly accessible in the time domain via pulsed light sources with pulse durations of the order of the dynamics of interest. Noteworthy in this regard are, among others, femtosecond optical lasers (OLs) and short-wavelength free-electron lasers (FELs) \cite{fels}, the latter enabling distinct element specificity due to the availability of short wavelengths. In fact, (X)FELs can provide a broad range of photon energies from the extreme ultraviolet (XUV) to hard X-ray pulses with femtosecond to attosecond duration and up to the mJ-level of pulse energy, thus opening a variety of opportunities to investigate nonlinear and ultrafast processes \cite{young2018roadmap, EmmaXFEL10, Hartmann18}.\\

\begin{figure*}[ht]
    \centering
    \includegraphics[width=\textwidth,width=0.8\linewidth]{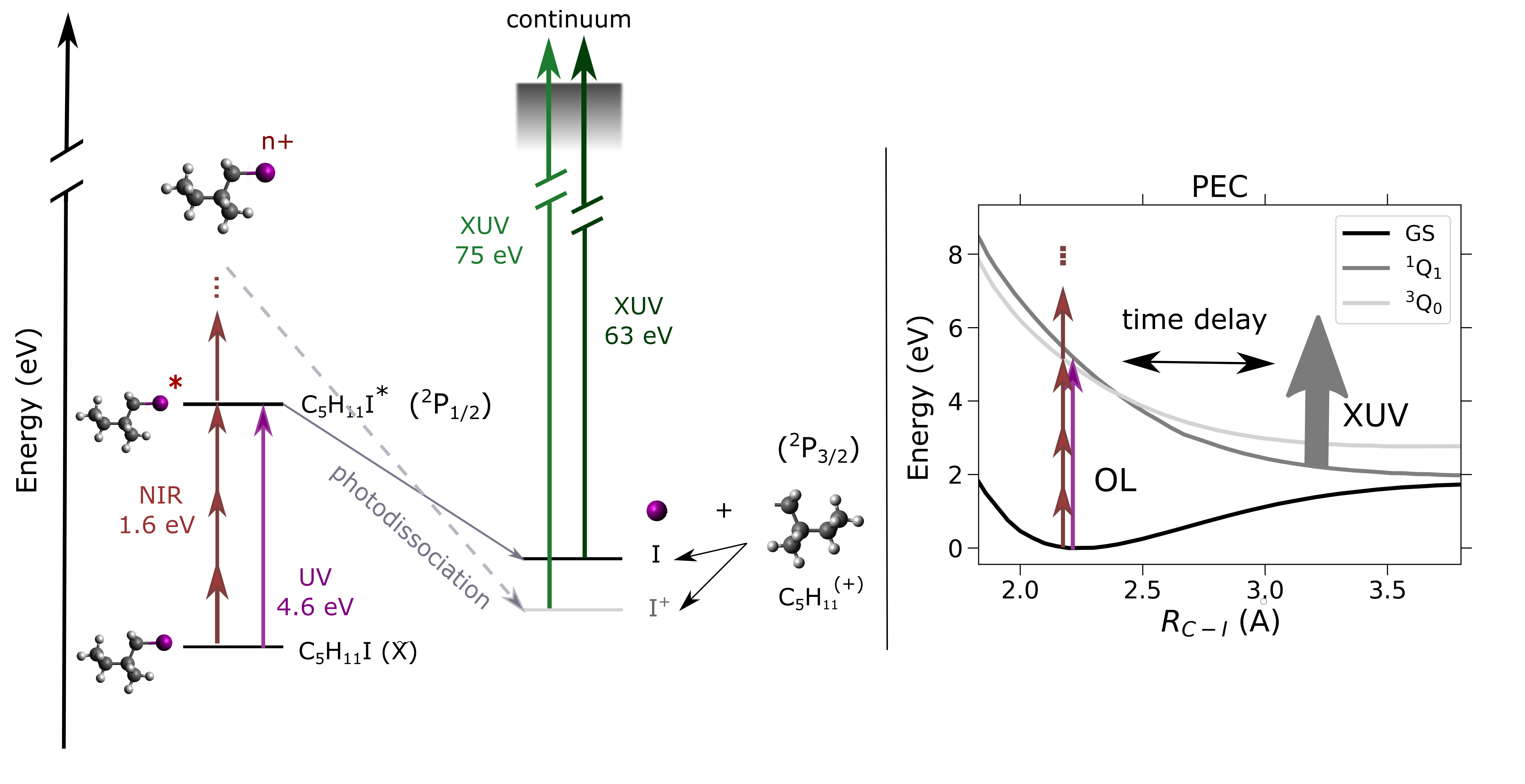}
    \caption{Schematic representation of the pump-probe experiment on 1-iodo-2-methyl-butane. The optical laser pulses photoexcite the molecule, either via one photon (4.6\thinspace eV, violet arrow) or multiphoton (1.6\thinspace eV, maroon arrows) absorption. Via the excited states, the molecule dissociates along the C-I bond involving either neutral or singly charged iodine which was subsequently probed via 4d photoionization using ultrashort XUV pulses. 
    The potential energy curves displayed on the right were calculated via $\Delta$CASSCF for the ground state and the two excited states $^1$Q$_1$ and $^3$Q$_0$ resulting from the neutral dissociation along the C-I bond. 
    }
    \label{fig:scheme}
\end{figure*}

Here, we report on the investigation of OL-induced fragmentation processes of isolated 1-iodo-2-methyl-butane molecules (C$_{5}$H$_{11}$I) resulting from interaction with ultraviolet (UV) or near-infrared (NIR) pulses with a wavelength (photon energy) of 267\thinspace nm (4.6\thinspace eV) and 800\thinspace nm (1.6\thinspace eV), respectively. The molecule as a prototypical chiral molecular system is interesting in several regards, such as its easy experimental accessibility and the presence of a heavy 'marker' atom outside its stereocenter as a possible observer, i.e. electron emission, site \cite{benjaminscience,motomura,mertens2016,koeckert,3-iodopyridine}. For both OL wavelengths, intensity-dependent ion yields have been obtained in order to identify different regimes of fragmentation into neutral or charged fragments. The respective changes have furthermore been investigated for their ultrafast time evolution using two XUV-FEL photon energies at 63 eV and 75 eV. These energies were chosen to predominantly ionize the 4d edges of the neutral and additionally singly-charged iodine in the fragments, respectively, resulting in similar electron kinetic energies (see figure \ref{fig:scheme}).  

Notably, when neutral iodine atoms are formed, the binding energies of the 4d electrons in iodine change from their values in the molecular environment $\approx$ 56.5\thinspace eV (4d$_{5/2}$) and $\approx$ 58\thinspace eV (4d$_{3/2}$), to isolated atomic iodine by $\approx +1.7$\thinspace eV \cite{felix}. For the fragmentation forming singly-charged iodine cations, we calculate a binding energy increase of the iodine 4d-electrons of about $12$\thinspace eV from the neutral molecule to the isolated iodine cation via $\Delta$CASSCF modeling (averaged over the two spin contributions) \cite{L1}. 
We establish selected fragmentation benchmarks for this prototypical chiral molecule with the future aim to investigate time-resolved photoelectron circular dichroism (PECD) phenomena \cite{ritchie,powis,boewering, ilchen2017, ilchen2021}, the latter being the reason for the choice of the same electron kinetic energies in neutral and ionic iodine.

%%%%%%%%%%%%%%%%%%%%%%%%%%%%%%%%%%%%%%%%%%
\bigskip
\section{Experiment}

The experiments were conducted at the BL1 beamline of the Free-Electron LASer in Hamburg (FLASH) \cite{Nature, feldhaus}. A double-sided velocity-map-imaging (VMI) spectrometer which is a part of the CAMP endstation \cite{J.Synch} was used to measure cations produced by photoionization of the chiral iodoalkane molecule C$_{5}$H$_{11}$I and its fragments, with a predominant scope of addressing the I 4d$_{3/2}$ and I 4d$_{5/2}$ vacancy states in different charge environments, thus involving different ionization potentials. At 63 eV the photoionization cross section of the iodine 4d is approximately 3\thinspace Mbarn \cite{crosssection1} and for 75 eV is approximately 13\thinspace Mbarn \cite{crosssection1, crosssection2}). These cross sections are similar to the sum of the photoionization cross sections of the remaining constituents.\\

In the center of the double-sided VMI spectrometer, the C$_{5}$H$_{11}$I molecular jet, discussed further below, was crossed with the UV or NIR pump laser, inducing the molecules' photo-excitation and/or -ionization. Both linearly polarized 'pump'-laser pulses were generated using an 800\thinspace nm Ti:Sapphire laser \cite{tisa} which propagates at a small angle of 1.5\thinspace $^\circ$ with respect to the FEL- and quasi-perpendicular to the molecular beam. The 800\thinspace nm (1.6\thinspace eV) fundamental wavelength was delivered at 10\thinspace Hz with an approximate pulse duration of 70\thinspace fs (FWHM). These pulses with a maximum energy of 1.8\thinspace mJ were focused to a spot size with a FWHM diameter on the order of 60\thinspace\textmu m $\pm ~20$\thinspace\textmu m. The 267\thinspace nm (4.6\thinspace eV) pulses were created by third-harmonic generation using a Beta Barium Borate (BBO) crystal \cite{bbo} with a maximum pulse energy of 176\thinspace\textmu J, a focus diameter of about 100\thinspace\textmu m $\pm  
\thinspace 40$\thinspace\textmu m, and a pulse duration of about 150\thinspace fs $ \pm~30$\thinspace fs. The following intensity ranges were covered by the UV and NIR pulses: $(0.4-5.2)\cdot10^{12}$\thinspace W/cm$^2$ and $(1.3-4.1)\cdot10^{14}$\thinspace W/cm$^2$. These values correspond to pulse energies between 0.05\thinspace\textmu J to 87.8\thinspace\textmu J and 0.36\thinspace mJ to 1.16\thinspace mJ, respectively.

The time-delayed circularly polarized \cite{vks} XUV pulses were generated by the FLASH1 FEL with a repetition rate of 10\thinspace Hz (single bunch mode) limited by the maximum achievable synchronization rate between OL and FEL under the given conditions. The pulse energies of the XUV pulses were chosen to be relatively low in order to minimize XUV-driven nonlinear effects on the sample and were 10\thinspace\textmu J for the setting using 63\thinspace eV and 20\thinspace\textmu J for 75\thinspace eV. They were focused into a spot with a diameter of about 10\thinspace\textmu m via Kirkpatrick-Baez optics, and the pulse duration was about 80\thinspace fs for both settings. The FEL focus size was chosen to be much smaller than the OL focus size in order to ensure optimum pump-probe contrast. The FEL pulse duration was retrieved via electron-beam diagnostics called 'LOLA' \cite{lola, lolaflash}. The pulse energy and timing fluctuations were recorded by the FLASH Gas Monitor Detector (GMD) \cite{GMD} and Bunch Arrival Monitor \cite{BAM}, respectively.\\

\begin{figure*}[ht]
\centering
\includegraphics[width=\textwidth,width=0.8\linewidth]{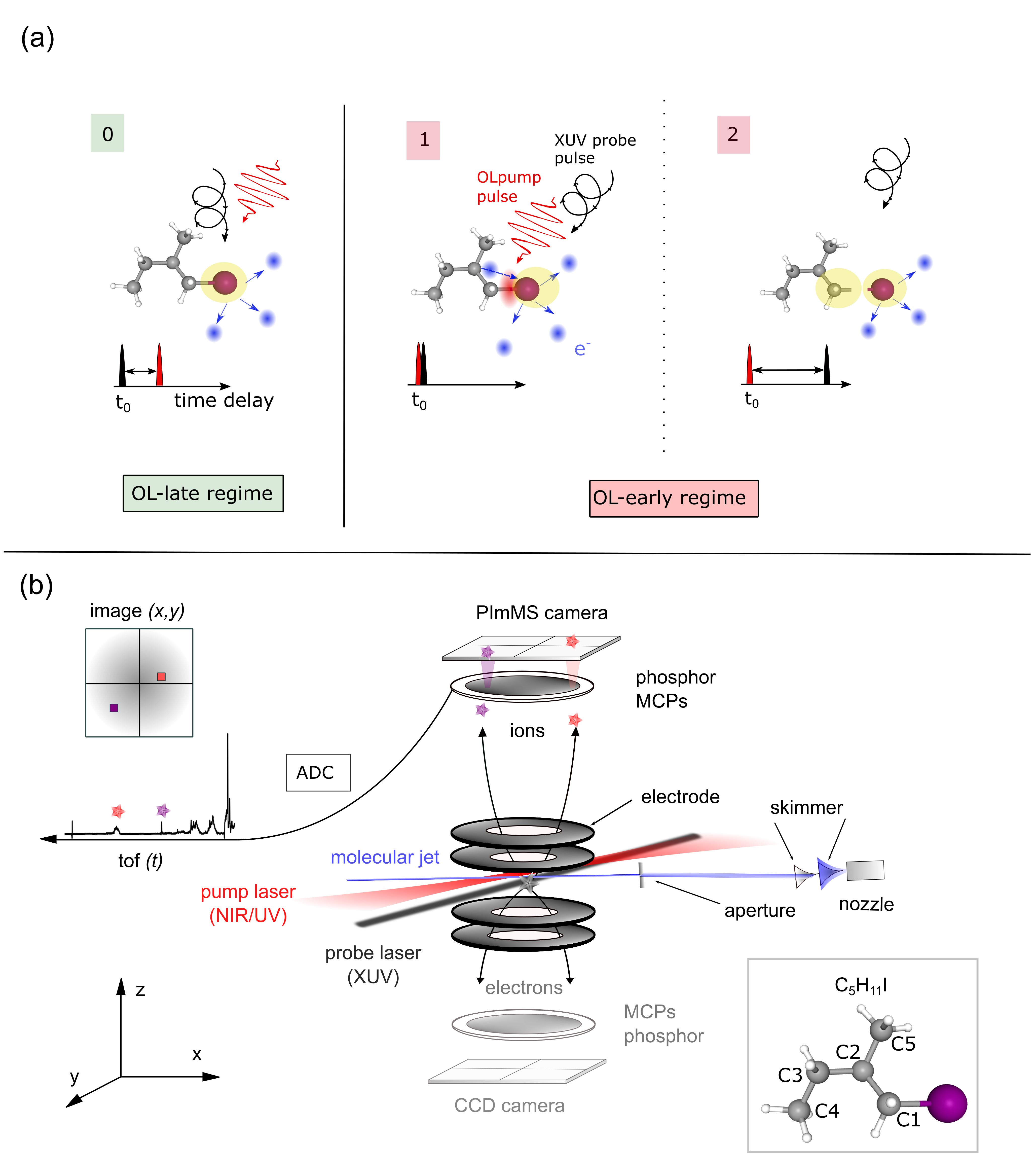}
\caption{(a) Sketch of molecular dissociation for two different regimes. The temporal overlap of the OL and XUV pulses is defined as time zero (t$_0$) \cite{felix}. In green, the OL-late regime is displayed. Here, the molecule is first illuminated by XUV and then by OL pulses (0). Secondly in red, the OL-early regime is displayed. Here, OL-pump pulses initiate a photodissociation of the molecule and time-delayed XUV pulses probe the iodine at two exemplary times (1) and (2). (b) Sketch of the experimental set-up: The injection of the molecules is depicted via the blue line system starting at the middle right. The skeletal of the molecule with annotated carbon atoms is visualized on the lower right. In the VMI spectrometer indicated by the round plates, the molecules are intersected by an optical pump laser (red) and the time-delayed FEL (black). The created charged particles, cations and electrons, were imaged on position-sensitive detectors (laboratory frame orientation indicated on the lower left). The ToF readout provides information about the different arrival times and thus mass/charge spectra. In the scope of this work, solely the cation data is analyzed and presented.}
\label{fig:exp}
\end{figure*}

The liquid molecular sample (99\% purity, Merck KGaA, Darmstadt, Germany and also synthesized in our laboratories in Kassel, Germany, according to established literature procedures \cite{chemie}) was stored in a cylinder and was heated up to 80\thinspace $^{\circ}$C with a positive gradient towards the jet nozzle in order to prevent clogging of the sample.  
The evaporated sample was seeded in a helium carrier gas at 1\thinspace bar backing pressure and transported into a heated sample-delivery line ending with a jet nozzle of a diameter of 30\thinspace\textmu m. The continuous molecular beam passed through two skimmers with orifice diameters of 150\thinspace\textmu m and 350\thinspace\textmu m, in this sequence. Finally, the molecular jet passes through an aperture of 4\thinspace mm opening for differential pumping, placed approximately 35\thinspace cm away from the interaction volume. As sketched in figure \ref{fig:exp}\thinspace(b), this doubly skimmed continuous supersonic molecular jet (in blue) was injected into the interaction region, and intersected by the OL (in red), and time-delayed XUV pulses of the FEL (in black). 

The voltages on the electrodes of the double-sided VMI were chosen such that charged particles with the same initial velocity were focused along the z-axis into the same point on the detector \cite{Eppink}. On both spectrometer sides, the particles hit 80\thinspace mm Chevron-stacked Multi-Channel Plates (MCPs) producing electron avalanches which were accelerated to P47 phosphor screens. For the positively-charged ions as subject of this work, a Pixel Imaging Mass Spectrometry (PImMS) \cite{PImMS} camera with 324 × 324 pixels was used as read-out of the screen illumination. The signals were furthermore independently monitored by a capacitive outcoupling of the MCP's current, thus providing time-of-flight information $(t)$ that can be converted to mass spectra. Contributions of stray-light-induced background electrons were reduced by fast HV switches (Behlke HTS31-GSM) that defined a temporal window of operation for the MCPs. Part of the electron data, concerning changes in the I 4d binding energy induced by single-photon 267 nm excitation, was published previously \cite{felix}.\\

Figure \ref{fig:exp}\thinspace (a) shows the two different time-delay regimes that were used within this work and are depicted as OL-late (in green) and OL-early (in red) regime. Time zero (t$_0$) defines the temporal overlap of the OL and XUV pulses. Firstly, in the OL-late regime in green (0) the molecule is illuminated by XUV pulses and afterward by the OL pulses. Secondly, in the OL-early regime, OL-pump pulses
initiate the photodissociation of the molecule. The time-delayed XUV pulses subsequently probe the iodine exemplarily at two different times \cite{pump-probe}: (1) XUV pulses arrive shortly after OL pulses. Here, the XUV pulses ionize the iodine and create charges, which can still be redistributed over the dissociating molecule \cite{benjaminscience,allum2023}. (2) At larger time delays between the two pulses, the molecule is dissociated, thus the charge transfer channel is closed. However, the Coulomb interaction between the fragments is still evolving. Ejected electrons are visualized in blue and the charge is indicated by a yellow surrounding. 

\bigskip
%\newpage
%%%%%%%%%%%%%%%%%%%%%%%%%%%%%%%%%%%%%%%%%%
\section{Results and Discussion}
\bigskip
\subsection{Intensity dependence of the pump process for UV and NIR}
The cation yields after dissociation of 1-iodo-2-methyl-butane induced by two OL wavelengths and at various OL intensities without XUV probe pulses are investigated.

\begin{figure*}[ht]
    \centering
    \includegraphics[width=\textwidth,width=13.5cm]
    {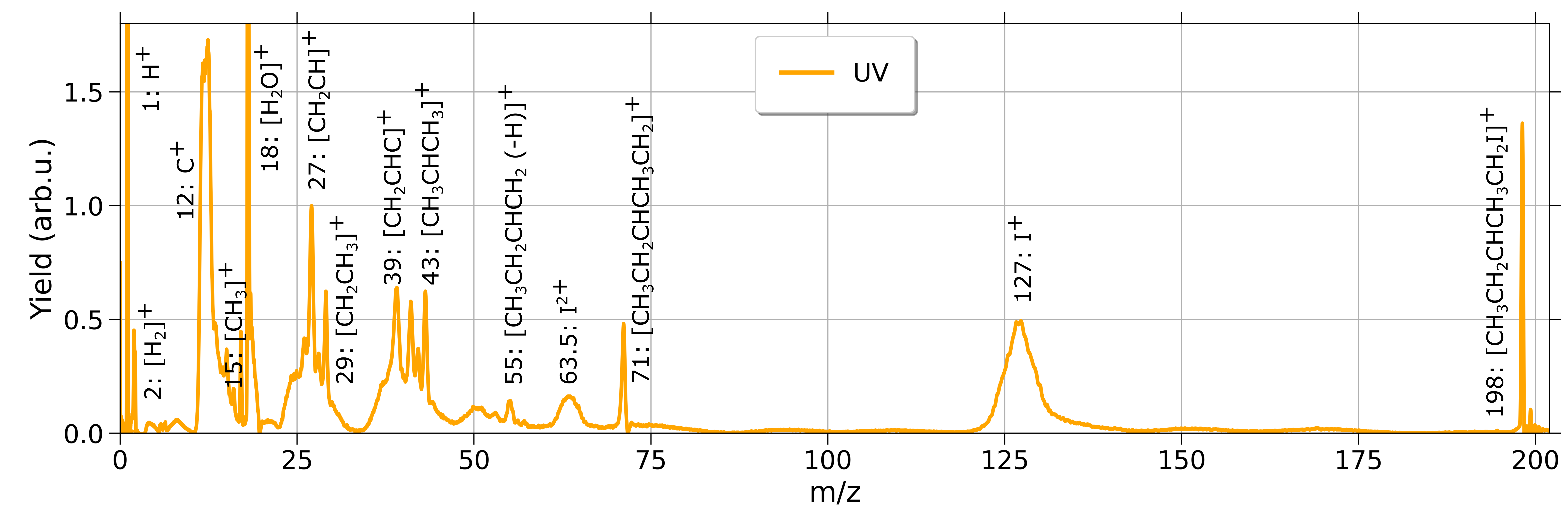}
    \caption{Mass spectrum for one exemplary UV (267\thinspace nm) laser intensity of 5.2$\cdot10^{12}$ W/cm$^2$. The most prominent peaks are labeled. The yield of the cations in arbitrary units, plotted on the y-axis and normalized to the peak $m/z$=27.
    }
    \label{fig:267}
\end{figure*}

\begin{figure*}[ht]
    \centering
    \includegraphics[width=\textwidth,width=13.5cm]
    {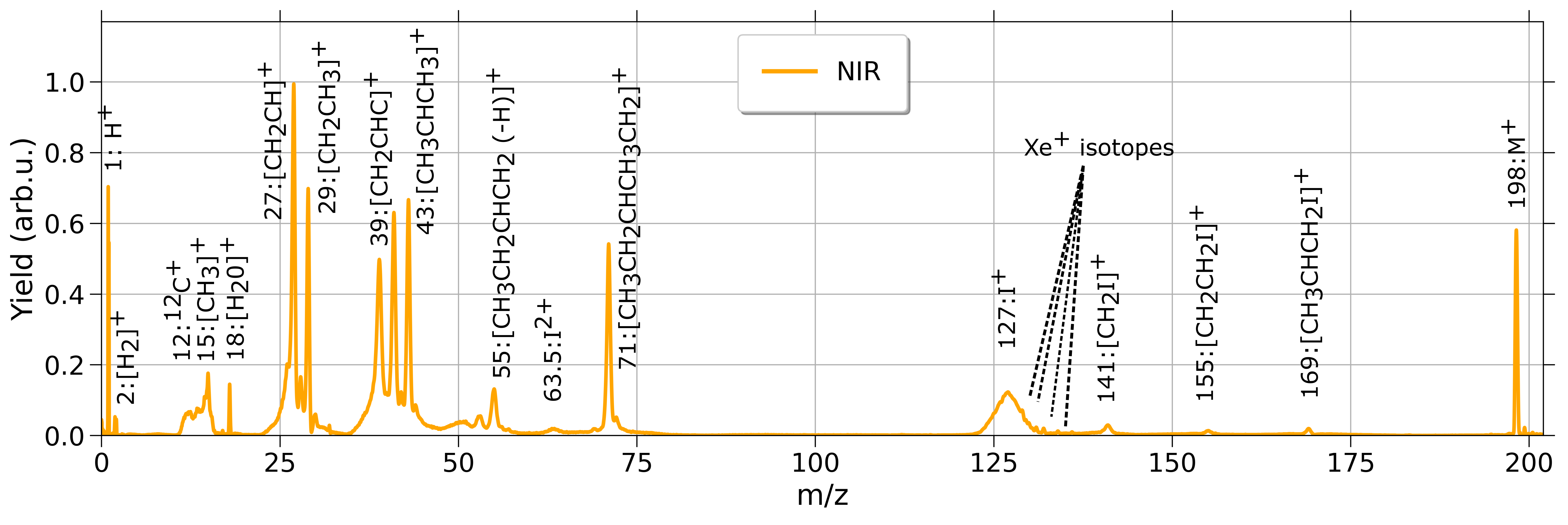}
    \caption{Mass spectrum for one exemplary NIR (800\thinspace nm) laser intensity 4.1$\cdot10^{14}$ W/cm$^2$. Plotted are mass over charges against their yield in arbitrary units, normalized to the highest peak $m/z$=27. 
    The iodine peak is partially overlapped by residual xenon, resulting from a calibration data set taken previously.}
    \label{fig:800}
\end{figure*}

The mass spectra obtained via UV or NIR pulses are illustrated in figures \ref{fig:267} and \ref{fig:800} for a selected exemplary intensity value, respectively. The spectra reveal distinctly different fragmentation patterns, as expected due to the differently triggered dissociation dynamics at different irradiation levels (see figure \ref{fig:scheme}). 


A variety of ionic fragments as well as the ionic intact molecule at $m/z$=198 are depicted to give an overview over exemplary fragmentation distributions.\\

In comparison, in the NIR mass spectrum (figure \ref{fig:800}) cations with the highest yield are the parent cation C${_5}$H${_{11}}$I${^+}$ at $m/z$=198, C${_5}$H${_{11}}$${^+}$ at $m/z$=71 and various alkyl groups C$_n$H$_x^+$ (n=2-4 and x=3, 5, and 7), which constitute the $m/z$ from 27 to 55.

Thus, for these NIR pulses, multiple fragmentation channels were populated, resulting in more ionic channels, and yielding several pathways with comparable abundance. Furthermore, the multiphoton dissociation induced by the NIR pulses yields three additional peaks which result from predominantly dissociating the molecule at a C-C and not the C-I bond. This can occur between the second and third carbon atom (CH$_3$CHCH$_2$I, $m/z$=169), the second and third plus second and fifth carbon atom (CH$_2$CH$_2$I$^+$, $m/z$=155) or the first and second carbon atom (CH$_2$I$^+$, $m/z$=141) (see illustration shown in the right bottom corner of figure \ref{fig:exp}\thinspace (b)). \\

\begin{figure*}[ht]
    \centering
    \includegraphics[width=\textwidth,width=13.5cm]{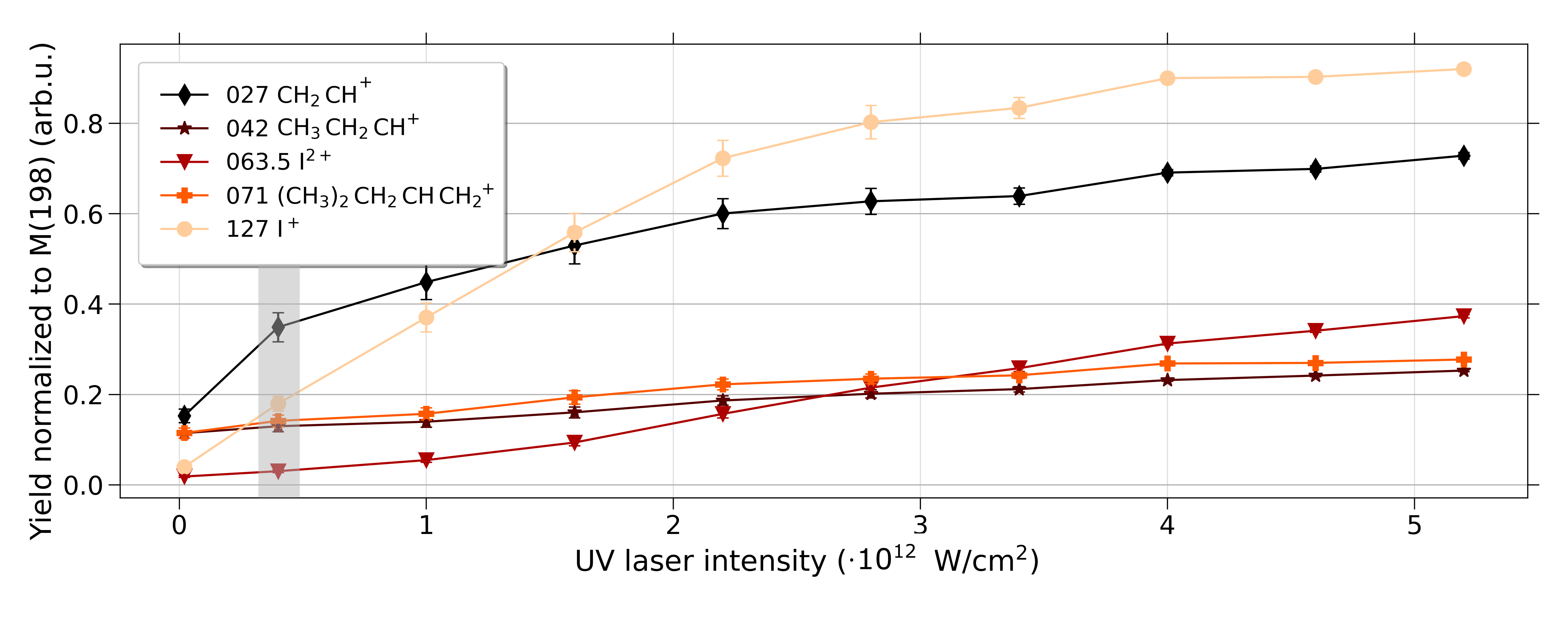}
    \caption{Yield of specific cationic fragments as a function of the UV laser intensity. Each value was normalized to the parent ion yield. Plotted are the mass over charges 27, 42, 63.5, 71 and 127. The lower intensity regime of the UV pulses, used for the below-presented UV-pump XUV-probe data, is shaded in light gray.} 
    \label{fig:267_ratio}
\end{figure*}

\begin{figure*}[ht]
    \centering
    \includegraphics[width=\textwidth,width=13.5cm]{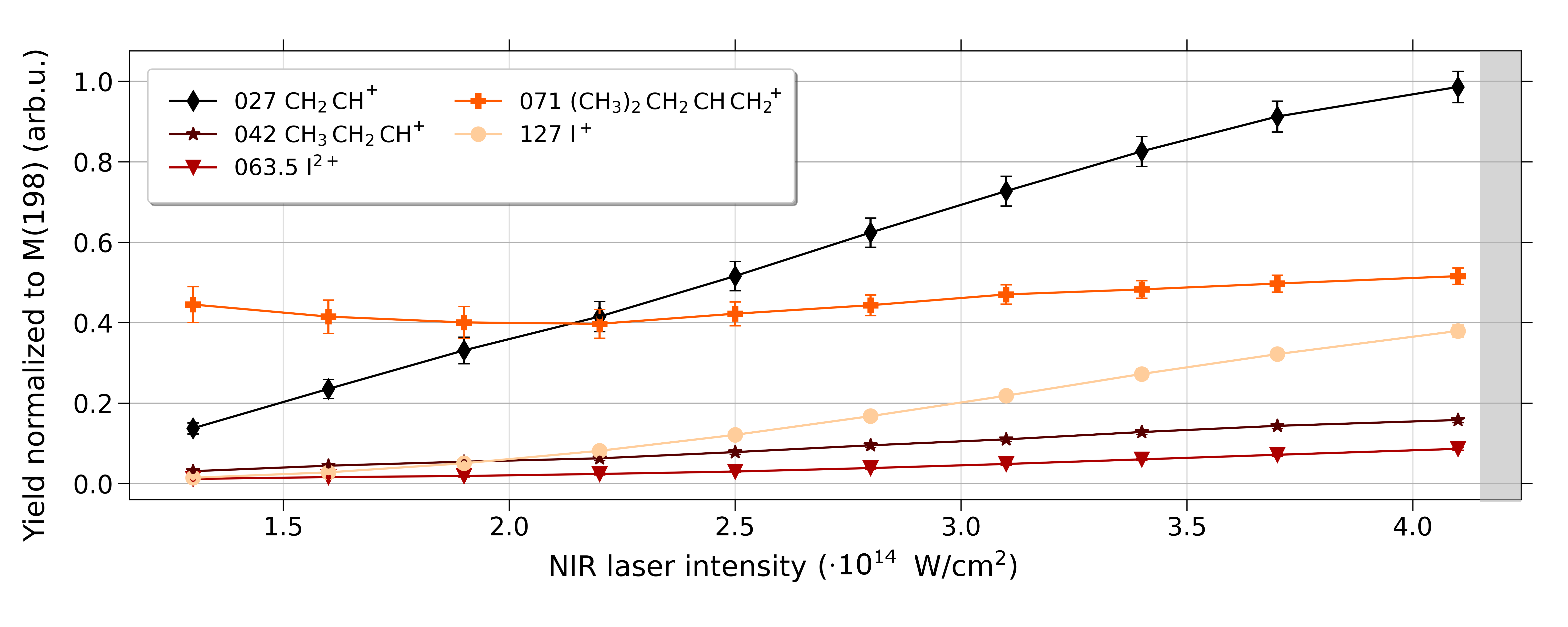}
    \caption{Yield of specific cationic fragments as a function of the NIR laser intensity. The yield is normalized to the yield of the parent cation with $m/z$=198. The higher intensity regime of the NIR pulses used, for the below-presented NIR-pump XUV-probe data, is shaded in light gray.}
    \label{fig:800_ratio}
\end{figure*}

Figures \ref{fig:267_ratio} and \ref{fig:800_ratio} depict the yield of selected cations as a function of the laser intensity. Each cation yield was normalized to the yield of the parent cation. Represented cations in the two figures are I$^+$ ($m/z$=127), C$_5$H$_{11}$$^+$ ($m/z$=71), I$^{2+}$ ($m/z$=63.5), C$_3$H$_7$$^+$ ($m/z$=42), and C$_2$H$_3$$^+$ ($m/z$=27). The changing yields, here with emphasis on the I$^{+}$, are an indication of different charge-up and dissociation pathways, which are the basis for the respective intensity choice for the time-resolving experiments presented below. The other fragments shown are a benchmark for reference. 


As shown in figure \ref{fig:267_ratio}, the yield of the singly-charged iodine fragment (light yellow dots) increases drastically as a function of the UV laser intensity and is quite dominantly produced at higher intensities. Since the goal of the UV-part of the experiment was to trigger neutral dissociation of the iodine atom the marked intensity slightly above the threshold of the appearance of the I$^{+}$ fragment was chosen in this case (UV intensity of 0.4$\cdot10^{12}$\thinspace W/cm$^2$, see gray area on the left of figure \ref{fig:267_ratio}) \cite{rebeccastructure, amini2018}. 

In contrast, as shown in figure \ref{fig:800_ratio}, the overall yield of the singly charged iodine in the investigated range of NIR intensities shows a more moderate increase towards higher intensities, and the total yield is smaller compared to other fragments. To ensure a high number of singly-charged iodine fragments for probing via the XUV pulses, a high NIR intensity of 4.3$\cdot10^{14}$\thinspace W/cm$^2$ was chosen to be used for the respective pump-probe experiments (see gray area on the right of figure \ref{fig:800_ratio}).

\subsection{Time-resolved results}
\label{time}
\bigskip
 
\begin{figure*}[ht]
    \centering
    \begin{minipage}[h!]{0.99\textwidth}
        \centering \includegraphics[width=\textwidth,width=13.5cm]{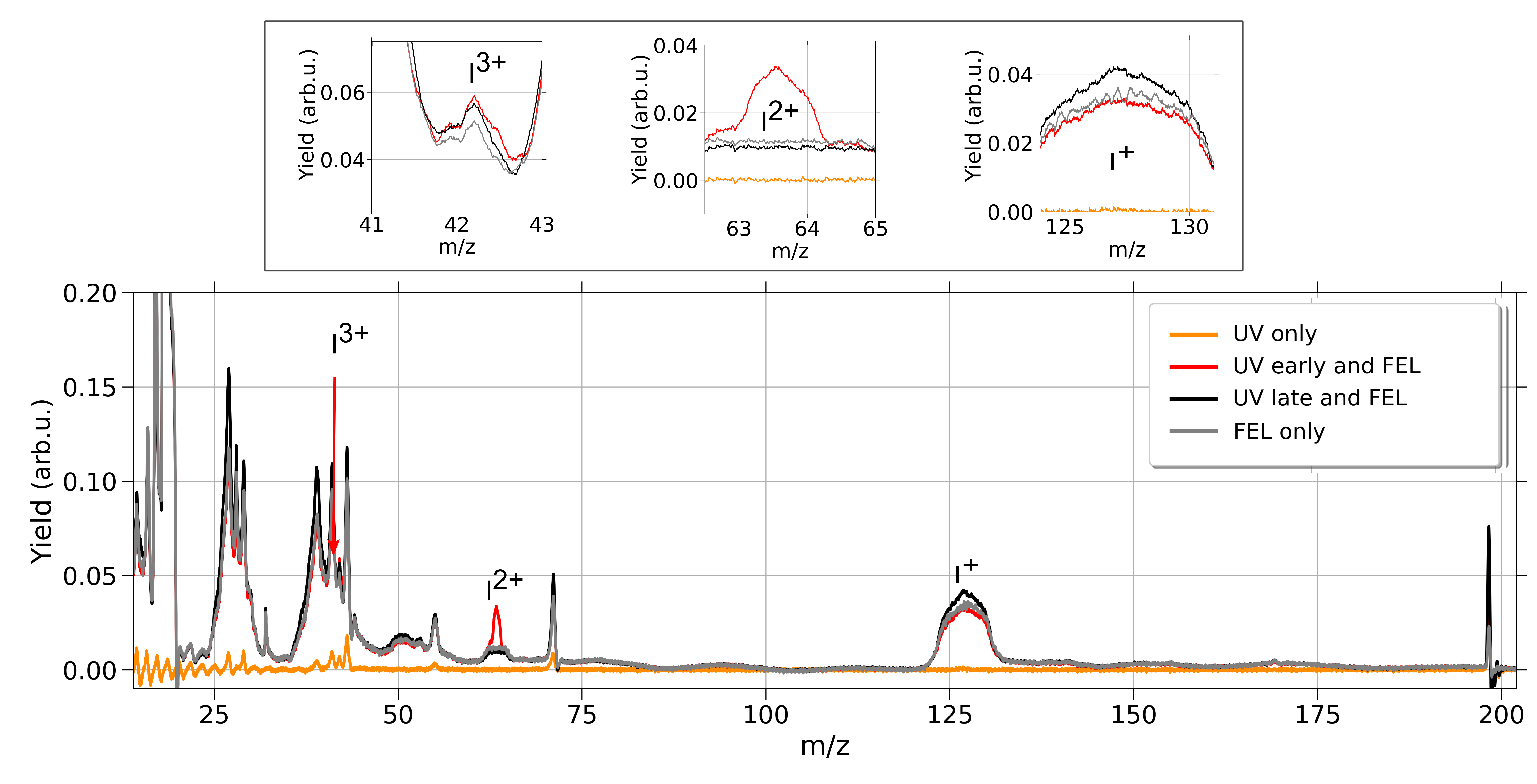}
        \small{(a)}
    \end{minipage}
    \hfill
    \begin{minipage}[h!]{0.99\textwidth}
        \centering \includegraphics[width=\textwidth,width=13.5cm]{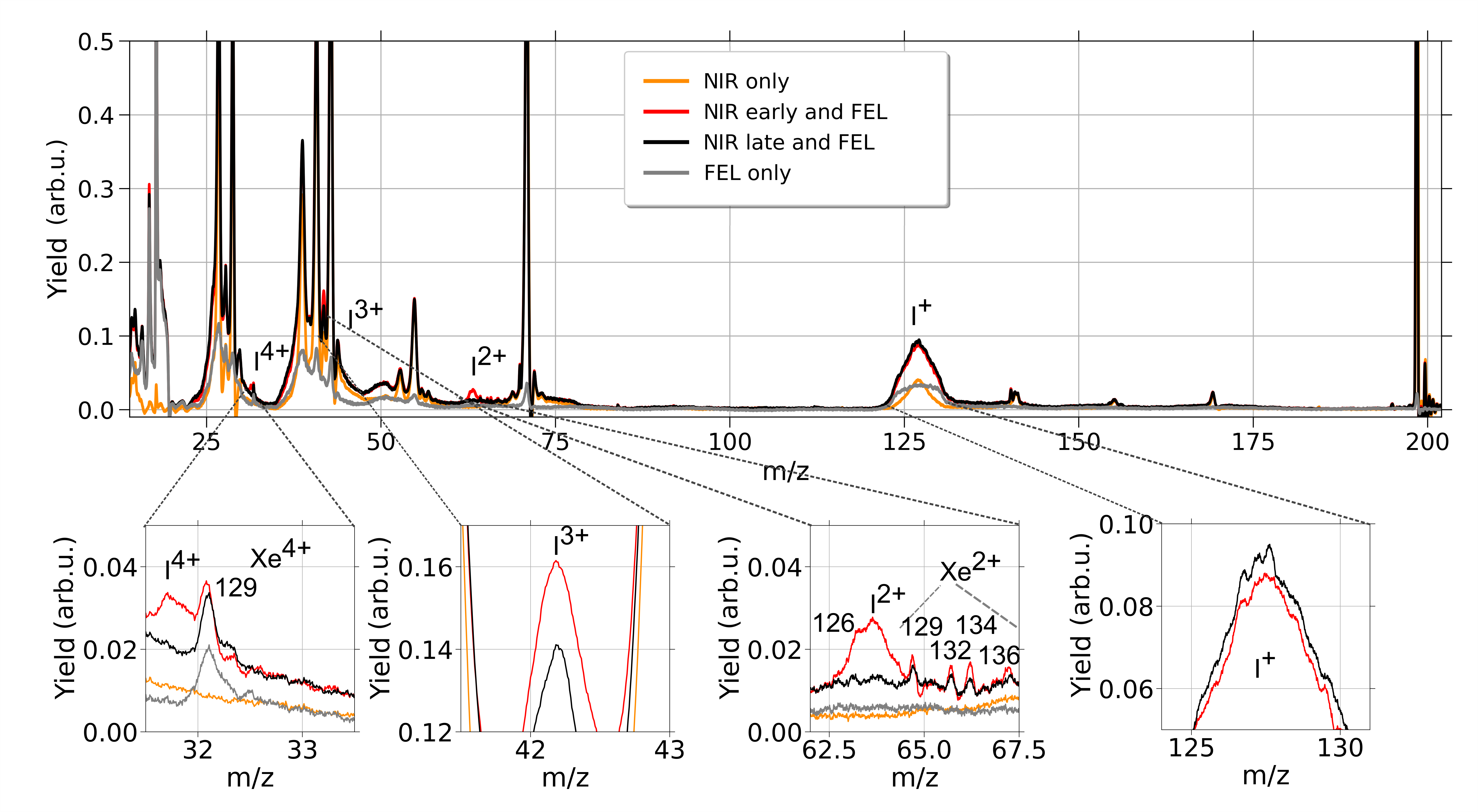}
        \small{(b)}
    \end{minipage} 
    \caption{Mass spectra of 1-iodo-2-methyl-butane for different pump and probe schemes, (a) for the UV and (b) for the NIR pulses with an intensity of 0.4$\cdot10^{12}$\thinspace W/cm$^2$ and 4.3$\cdot10^{14}$\thinspace W/cm$^2$, respectively. Compared are mass spectra resulting from the OL (orange), from the XUV-FEL ionization (gray), from OL-pump pulses 500\thinspace fs (UV) / 2\thinspace ps (NIR) before the XUV-probe pulses (red, OL-early regime) and lastly from the probe pulses 500\thinspace fs (UV) / 1\thinspace ps (NIR) before the pump pulses (black, OL-late regime). Enlarged views are plotted for (I$^{4+}$), I$^{3+}$, I$^{2+}$, and I$^{+}$ fragments. Short cation flight times are subject to ringing due to electronic feedback from the pulsing of the high voltage of the electron detector. Mass-to-charge ratios <$m/z$=13 are therefore not analyzed. Xenon peaks result from residual gas of a previous calibration measurement.} 
    \label{fig:late_early}
\end{figure*}

The goal of the time-resolving investigations was to distinguish between Coulomb-driven and neutrally dissociative fragmentation for NIR-pump and UV-pump experiments, respectively, at specific intensity choices based on the previous findings to probe the timescales over which these dynamics occur. For the case of UV-pump XUV-probe (h$\nu$=63\thinspace eV), we put specific emphasis on an enhanced level of neutral dissociation via the excited states $^3{Q_0}$ and $^1{Q_1}$ \cite{neutraldiss} by using a moderate UV intensity of 0.4$\cdot10^{12}$\thinspace W/cm$^2$. In contrast, for the NIR-pump XUV-probe (h$\nu$=75\thinspace eV) case, a relatively high NIR intensity of 4.3$\cdot10^{14}$\thinspace W/cm$^2$ was employed in order to ensure a high number of singly-charged iodine fragments that can be probed by the XUV pulses. Note that this value lies even slightly beyond the presented laser-intensity scan in figure \ref{fig:800_ratio}.
Figure \ref{fig:late_early} shows mass spectra resulting from the two-color experiments using either OL-laser pulses only (orange) or XUV pulses only (gray) as well as laser-early and laser-late cases (red and black, respectively). For the mass spectra involving XUV pulses, the pulse energy fluctuation was corrected on a shot-to-shot basis via the GMD data.

The UV-pump XUV-probe mass spectra, presented in figure~\ref{fig:late_early}\thinspace (a) show that the total yield of the ionic fragments stemming from UV ionization (depicted in orange) is much lower compared to that generated by XUV pulses (shown in gray). Notably, and consistent with the deliberately adjusted low-intensity UV pulses, almost no charge states of iodine and only a limited number of ionic channels, primarily encompassing the parent, C$_5$H$_{11}$, and alkyl groups, are generated. Note that for low $m/z$ values, ringing of the Behlke switches overlaps the mass spectra. The XUV photons with an energy of 63\thinspace eV 

lead to a significant formation of I$^{+}$, I$^{2+}$, and I$^{3+}$ even at the employed low intensity. Since this photon energy isn't sufficient for the core ionization of ionic iodine, neither within the molecule nor as an atomic cation, double Auger-Meitner relaxation is expected to be the most prominent origin for these contributions \cite{doppelauger}, but also excited states in the cation and/or charge transfer mechanisms from other sites of the molecular system can contribute to these charge states. As can be seen in figure~\ref{fig:late_early}\thinspace (a), for the UV-early case, the yield of singly-charged iodine is significantly lower than for the UV-late case, which indicates that the FEL produces excited states that allow the UV to finally ionize the iodine.

Pump-probe features can be seen for I$^{2+}$ in the mass spectrum for the UV-early case (see red mass spectrum in figure \ref{fig:late_early}\thinspace (a)). This signal predominantly results from the discussed neutral one-photon dissociation of the molecule \cite{neutral1, neutral2, neutral3, neutral4} into iodine and the alkyl group.  This is followed by the ionization of the iodine atom via the XUV pulses and a subsequent Auger-Meitner decay which leads to the creation of I$^{2+}$ with I$^*$ in spin-orbit excited states:

\begin{equation}
\begin{split}
\rm{C_5H_{11}I + UV \rightarrow C_5H_{11} + I^{(*)}} \\
\rm{C_5H_{11} + I^{(*)} + XUV \rightarrow C_5H_{11} + I^{2+}}
\end{split}
\label{eq:region3uv}
\end{equation}

\noindent At larger internuclear distances, the charge transfer between the fragments becomes impossible, and the charge vacancy cannot be distributed between the atomic iodine and C$_5$H$_{11}$, thus the evolution of the time-dependent I$^{2+}$ yield can be an indication for the closing of the charge-transfer channel \cite{benjaminscience}. The same effect can be identified for the delay-dependent peak at $m/z$=42.3, representing the I$^{3+}$ fragment, as expected from a double-Auger-Meitner yield in the order of 15$\%$ \cite{doppelauger, Forbes2020}. 

The mass spectra represented by the black lines display molecules that first interact with the XUV pulses and then with the 500\thinspace fs delayed UV pulses (OL-late). The UV pulses appear to ionize a variety of excited molecular fragments generated by the initial XUV pulses, as it was concluded for the iodine above.\\ 

For the NIR-laser pulses the pump-probe mass spectra shown in figure \ref{fig:late_early}\thinspace (b) are fundamentally different, mainly due to the much higher intensity of the NIR pump-laser pulses, and their inability to resonantly excite the molecule through single-photon absorption. The multiphoton dissociative ionization results in several ionic fragments (orange), as well the iodine charge state I$^{+}$. 'NIR-early' peaks can be seen in the mass spectrum for the case of a 2\thinspace ps delay (shown in red). Compared to the previously discussed UV-case the same peaks at $m/z$=63.5 (I$^{2+}$) and $m/z$=42.3 (I$^{3+}$) can be identified, with similar abundance for the former and an increased strength for the latter. I$^{2+}$ fragments appear less prominent, due to the lower number of neutral iodine created via the relatively strong NIR pulses (see also discussion below for the time-resolving scans over a longer delay range with multiple steps). In contrast, the additional I$^{+}$ fragments created by the NIR pulses, lead to a more prominent pump-probe signal at the I$^{3+}$ fragment as a consequence of the higher XUV photon energy in this part of the experiment, which is sufficient for core-ionizing singly charged iodine. The observed pump-probe signal in the I$^{4+}$ yield is consistent with the mentioned double-Auger yield. Residual Xe from a previous calibration is annotated in the inset, also showing a small pump-probe effect as can be expected from similar mechanisms. 

\begin{figure*}[ht]
    \centering
    \begin{minipage}[t]{0.9\textwidth}
        \centering \includegraphics[width=\textwidth,width=13.5cm]{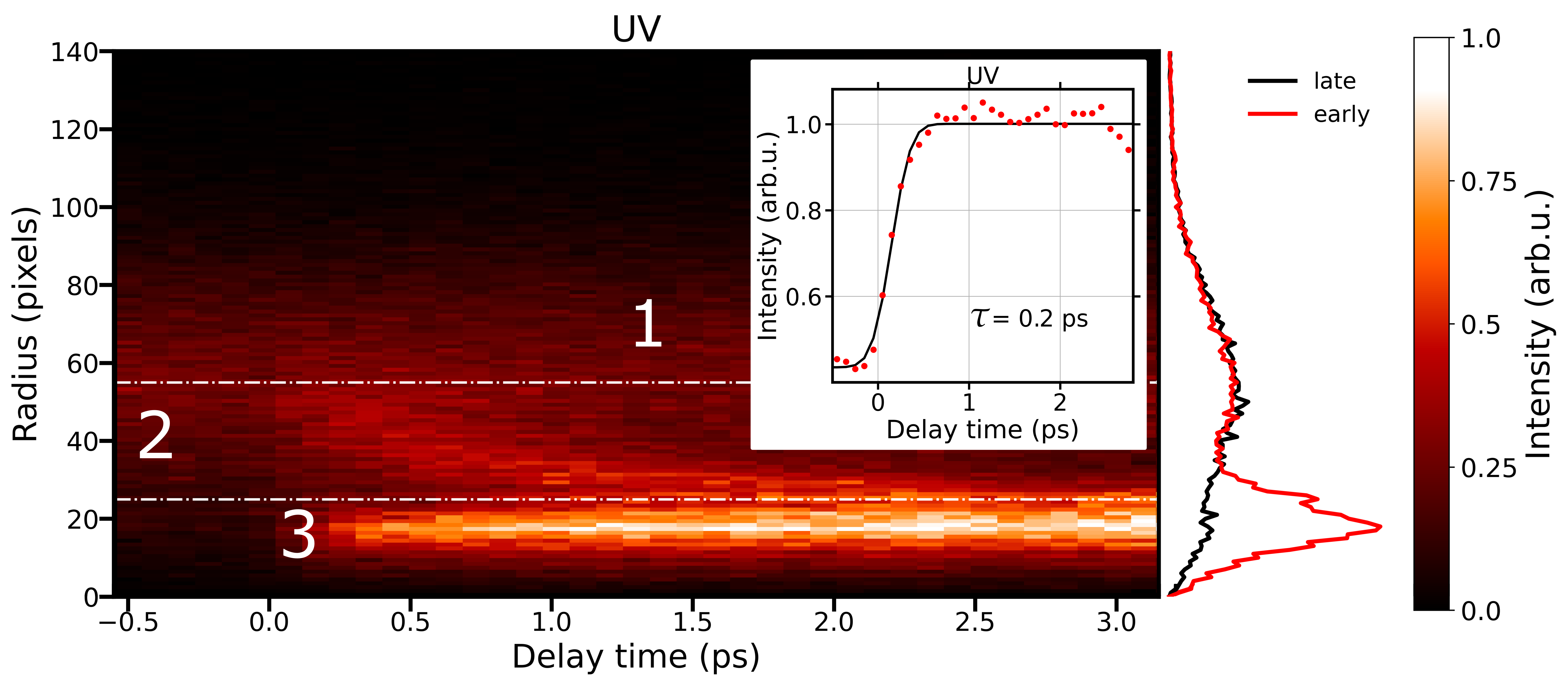}
        \small{(a)}
    \end{minipage}
    \hfill
    \begin{minipage}[t]{0.99\textwidth}
        \centering \includegraphics[width=\textwidth,width=13.5cm]{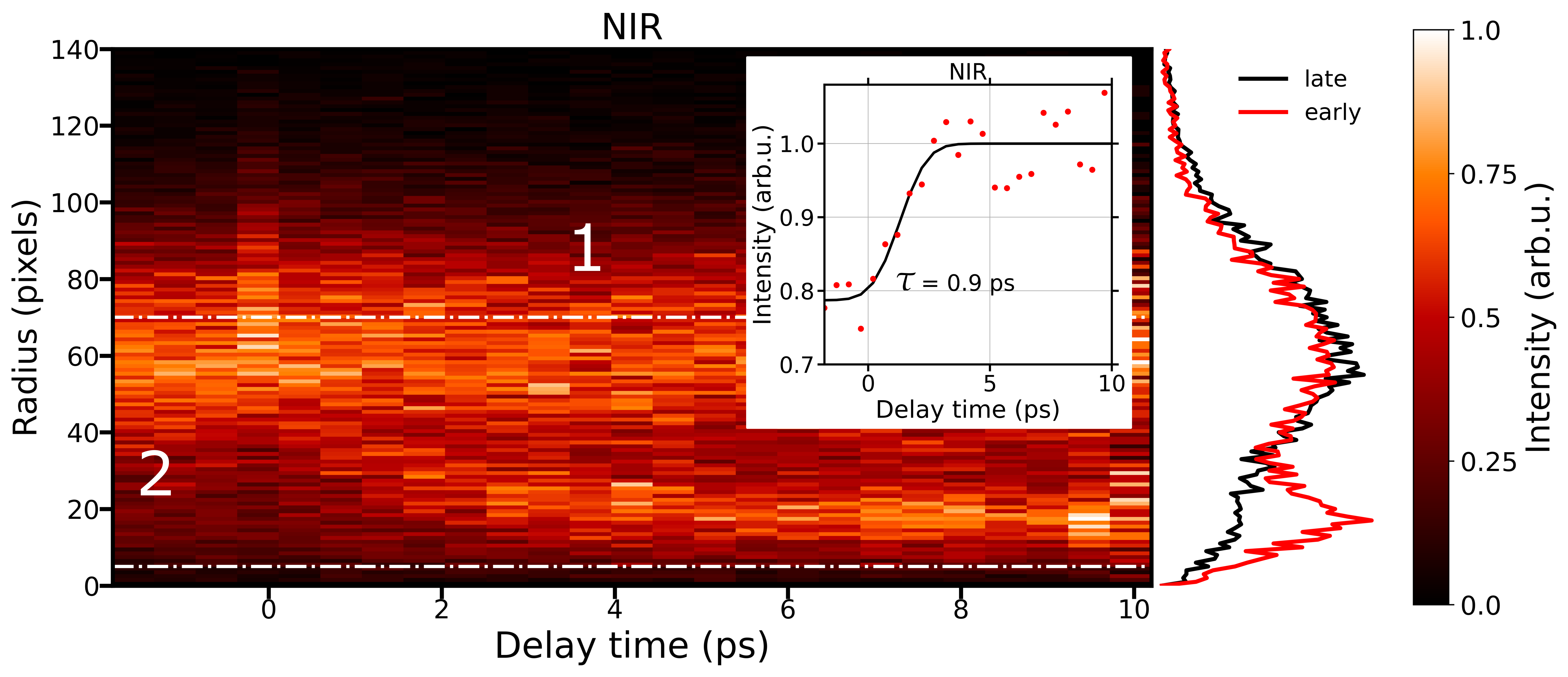}
        \small{(b)}
    \end{minipage}
    \caption{Radial distribution maps for the delay-dependent kinetic-energy distributions of I$^{2+}$: (a) UV pump - XUV probe and (b) NIR pump - XUV probe. The projections of the radial distributions are displayed next to the maps: UV/NIR late in black for delays -0.25/-1\thinspace ps and UV/NIR early in red for delays +2/+5\thinspace ps. Three regions are marked, named 1, 2 and 3. The insets represent the integrated yield of the ion-energy channel I$^{2+}$  within the dashed-dotted white lines capturing region 2 as a function of pump-probe delay (red points) with a fit to a normal cumulative distribution function (black line). The central value of the fitted Gaussian function is indicated as $\tau$.}
    \label{fig:heatmap}
\end{figure*}

Delay-dependent effects can be studied in more detail by investigating the kinetic energy (KE) of the relevant ionic fragments such as I$^{2+}$ as a function of pump-probe delay, see figure \ref{fig:heatmap}\thinspace (a) for UV and (b) for NIR pulses. For the UV-induced fragmentation, a scan in steps of 100\thinspace fs, covering a 3.5\thinspace ps range starting at -0.5\thinspace ps and ending at 3 ps, is presented. The covered delay range for the NIR-case was 12\thinspace ps, ranging from -2\thinspace ps to 10\thinspace ps, with scan steps of 500\thinspace fs since the observable processes cover a longer range than in the UV-case. Positive delays correspond to the OL-early and negative delays to OL-late regime. 

A projection of the kinetic-energy dependent yields is plotted for the OL-early regime (red) and one for the OL-late (black) regime vertically between the radial distribution map and intensity color bar. The y-axis represents the radial distribution in pixels of the measured I$^{2+}$ fragments, which was derived from the velocity-map images from the PImMS camera. The radius is proportional to the magnitude of the initial ion velocity in the detector plane \cite{Eppink}. This representation of the data in Figure \ref{fig:heatmap} was chosen due to inability to perform high-quality inverse Abel transformation of the data (to yield the underlying three-dimensional velocity distribution) due to lack of statistics, particularly in the NIR-pump XUV-probe data.

In these delay-dependent ion velocity distributions, three distinct regions of signal can be identified, which were marked as Region 1, 2 and 3. Region 1 comprises a broad, high velocity ($\approx$30 to 100\thinspace pixels, corresponding to KEs between 1 and 3\thinspace eV),  feature which is present over all pump-probe delays. This feature originates from XUV-only Coulomb explosion of the molecule to produce mutually repelling I$^{2+}$ and ionic alkyl cofragment(s). Given the propensity for single-photon XUV ionization to yield a doubly charged cation (which cannot Coulomb explode into I$^{2+}$ and other charged fragments), it is believed that much of this Coulomb explosion signal arises from absorption of multiple XUV photons. This process is more likely for the experiments with the NIR-pump pulse, owing to the higher XUV intensity employed.

Two time-dependent features are shown in Figure \ref{fig:heatmap}, labeled 2 and 3 respectively, both of which appear shortly after time-zero. Time-zero is defined according to the method discussed in detail in \cite{felix}. The former feature has a delay-dependent KE, decreasing at longer pump-probe delay. In contrast, Region 3 comprises signal with a constant low KE. As observed in previous OL-pump XUV-probe experiments, Region 2 is indicative of a Coulomb repulsion between fragments that is prompted by the XUV probe pulse \cite{benjaminscience, rebeccastructure,amini2018,allum2020,allum2023}. At longer pump-probe delays, the fragments are at greater separations, and thus this Coulomb repulsion decreases. Region 3, which is only clearly observed in the case of UV photoexcitation, arises when I$^{2+}$ is produced by the XUV pulse in the absence of any charged co-fragments. This is indicative of a neutral photodissociation prompted by the optical laser, followed by XUV-ionization solely at the dissociated, isolated, iodine atom (see also equation \ref{eq:region3uv}). The KE distribution of the detected I$^{2+}$ ions (with a mean KE of $\approx$0.25~eV) reflects this photodissociation process to predominantly yield spin-orbit excited I* photoproducts, and is consistent with previous measurements \cite{neutraldiss,felix}. In the case of NIR excitation, Region 3 signal could arise from a multi-photon induced neutral photodissociation to yield C$_5$H$_{11}$+I followed by 4d ionization of the I photoproduct or from dissociative ionization to C$_5$H$_{11}$+I$^+$ followed by valence ionization of the I$^+$ photoproduct. The lack of a clear Region 3 signal in the NIR experiment reflects the low likelihood of these processes.

As mentioned above, the delay-dependent Coulomb repulsion contribution to the Region 2 signal indicates two mutually repelling charges following interaction with the XUV-probe pulse. In the case of UV photoexcitation, which is believed to predominantly induce single-photon neutral dissociation under the experimental conditions, this Coulomb repulsion can be induced by XUV-based valence ionization of the C$_5$H$_{11}$ fragment alongside 4d ionization of the neutrally dissociated iodine (see equation \ref{eq:region2uv}). 

\begin{equation}
\begin{split}
\rm{C_5H_{11}I + UV \rightarrow C_5H_{11} + I^{(*)}} \\
\rm{C_5H_{11} + I^{(*)} + XUV \rightarrow C_5H_{11}^{+} + I^{2+}}
\end{split}
\label{eq:region2uv}
\end{equation}

\noindent Minor contributions to this signal may also arise from multiphoton dissociative ionization to yield a charged alkyl fragment and a neutral iodine atom which is subsequently core ionized. \\

In the case of NIR excitation, which predominantly photoionizes the molecule under the given conditions, this signal is believed to mainly originate from such dissociative ionization processes (see equation \ref{eq:region2nir}). 

\begin{equation}
\begin{split}
\rm{C_5H_{11}I + NIR \rightarrow C_5H_{11}^{+} + I}\\ 
\rm{C_5H_{11}^{+} + I + XUV \rightarrow C_5H_{11}^{+} + I^{2+}}
\end{split}
\label{eq:region2nir}
\end{equation}

\noindent We note that a detailed modeling of the time-evolving KE distribution associated with Region 2 is challenging in the present experiment due to uncertainty in the identity of the charged alkyl cofragment and the trajectories of any neutral fragments produced simultaneously. As seen in the mass spectra reported in figure \ref{fig:late_early}, the NIR pulse produces a range of alkyl ions. Whilst UV excitation will yield predominantly C$_5$H$_{11}$ fragments, the subsequent XUV ionization may yield smaller daughter ions. 

To examine the timescales of the underlying processes leading to Region 2 signal in the two excitation regimes, the integrated ion intensities of velocity ranges of interest are shown as a function of pump-probe delay in the insets of Figure \ref{fig:heatmap}\thinspace(a) and (b). To quantify these timescales, the delay-dependent intensities were fitted with a Gaussian cumulative distribution. In the case of UV excitation, this signal rises on a timescale of a few hundred femtoseconds ($\tau=200$~fs), close to the estimated temporal resolution of the experiment. In the case of UV excitation, a velocity range was chosen which excludes Region 3. A similar rise time as in earlier studies is observed for the lower velocity, Region 3 signal \cite{felix}. In contrast, for NIR excitation, the timescale for the formation of Region 2 signal is several times slower ($\tau=900$~fs), implying that there is a substantial time between interaction with the NIR pulse and the ultimate fragmentation. 

\begin{figure}[p]
    \centering
    \begin{minipage}[t]{0.4\textwidth}
        \centering \includegraphics[width=\textwidth,width=\textwidth]{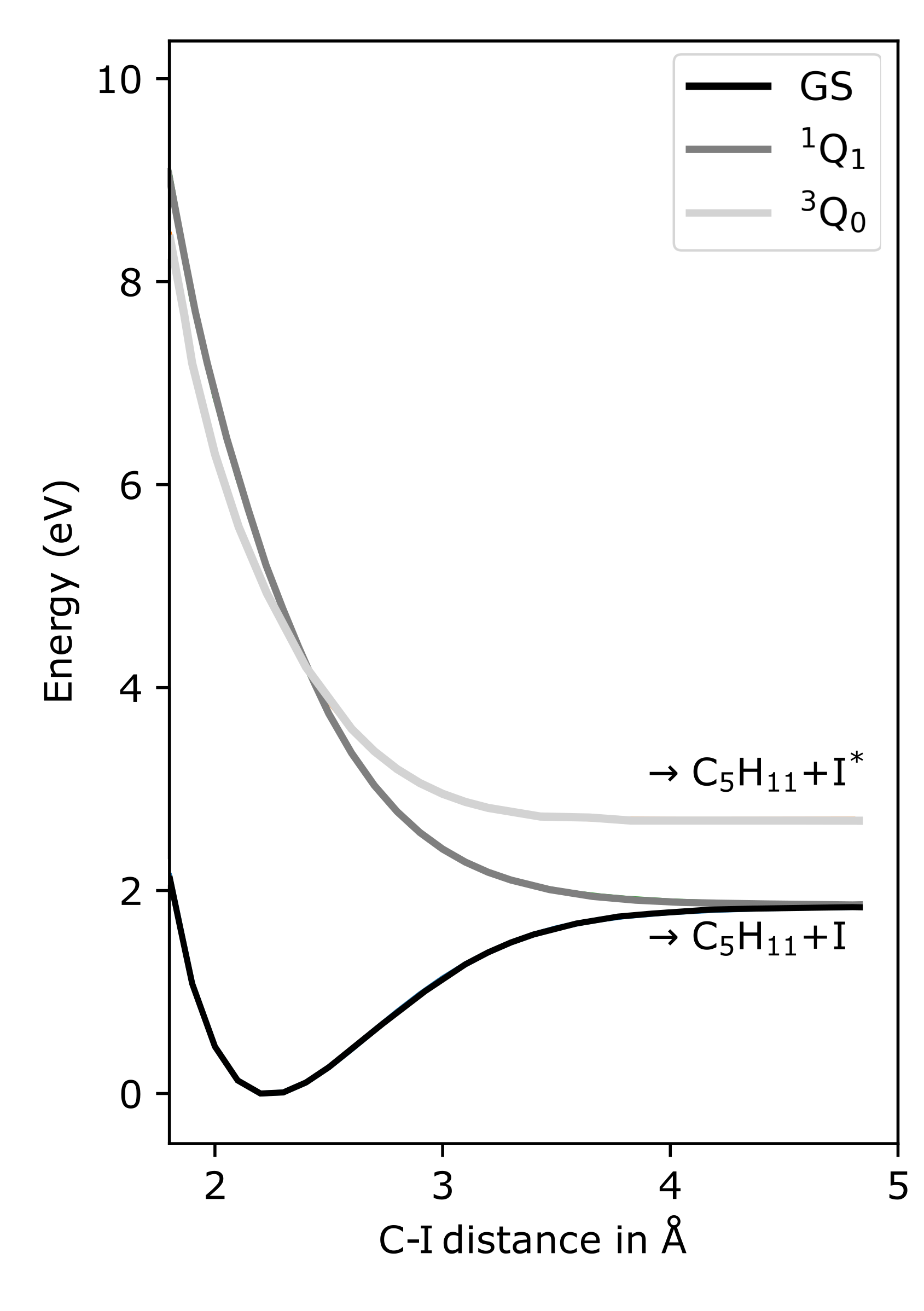}
        \small{(a)}
    \end{minipage}
    %\hfill
    \begin{minipage}[t]{0.353\textwidth}
        \centering \includegraphics[width=\textwidth,width=\textwidth]{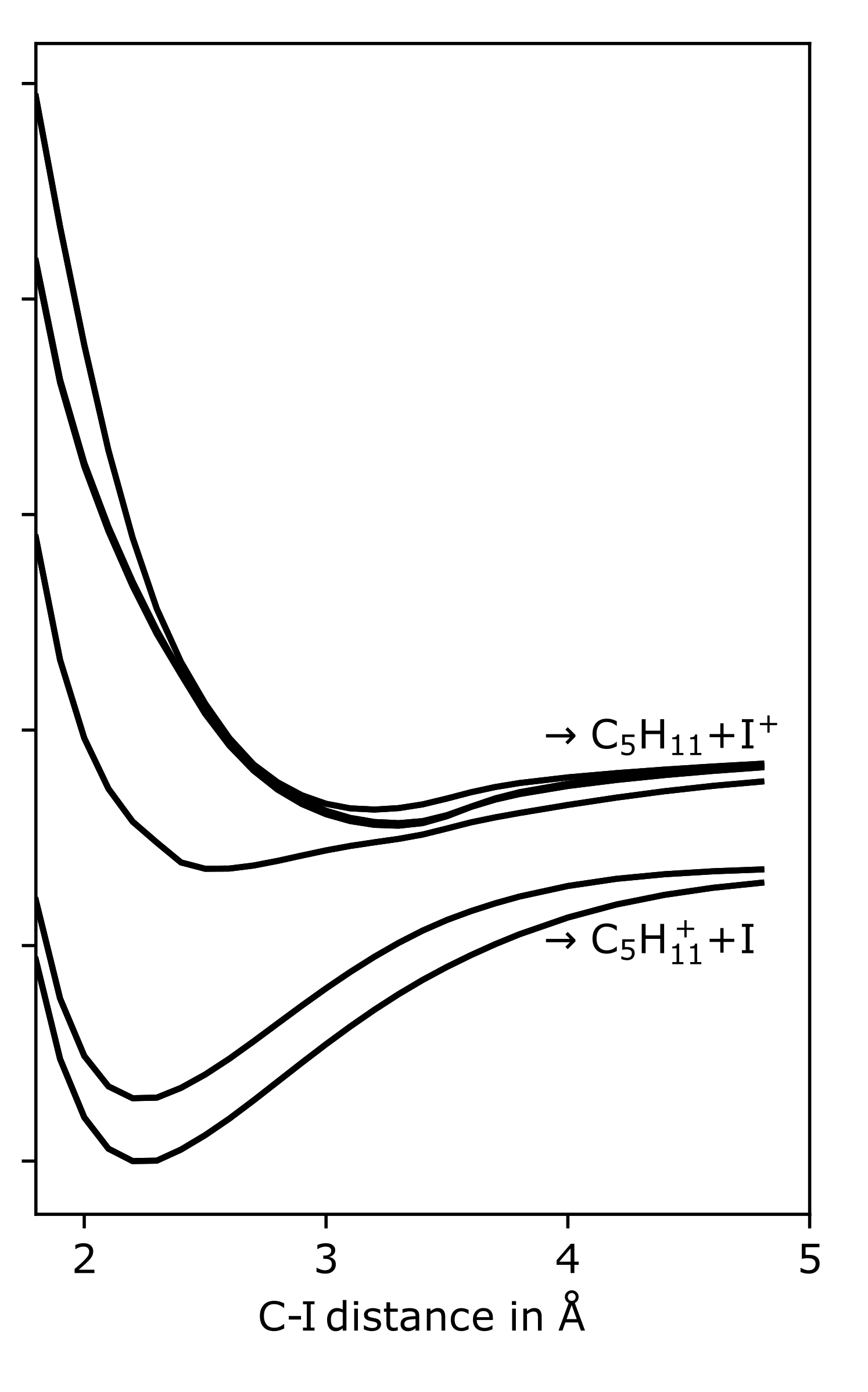}
        \small{(b)}
    \end{minipage}
    \caption{(a) Potential energy curves along the C-I bond for the ground state and the relevant $^3{Q_0}$ and $^1{Q_1}$ excited states.
    (b) Potential energy curves along the C-I bond for the energetically
    lowest states of the cation.}
    \label{fig:PEC}
\end{figure}

The calculated potential energy curves (PECs) shown in figure \ref{fig:PEC} suggest a reason for the strong difference in time scales. For the neutral molecule, a state-averaged complete active space (SA-CASSCF) calculation was conducted using an orbital space of four orbitals with six electrons and averaging over the three lowest singlet states employing the 6-311G(d,p) basis set \cite{L1,L2}. From the obtained set of orbitals, the basis states for the spin-orbit coupling (SOC) calculations were constructed consisting of the six lowest triplet and four lowest singlet states were obtained by diagonalizing the configuration interaction (CI) matrix in this active space. The Breit–Pauli Hamiltonian was then diagonalized in these basis states. The PECs of the $^3{Q_0}$ and $^1{Q_1}$ excited states of the neutral molecule, which are populated by the UV pulses, are dissociative along the C-I coordinate (figure \ref{fig:PEC}\thinspace (a)) and lead to very prompt photodissociation.

For the molecular cation mainly addressed by NIR pump pulses, a SA-CASSCF calculation involving 6 orbitals and 5 electrons was performed, state averaged over 6 doublet states, and employing the same atomic orbital basis set as for the neutral case. The Breit–Pauli Hamiltonian was diagonalized considering the 4 lowest quadruplet and 6 lowest doublet states. The calculations were performed with Molpro version 2020.1 \cite{L3} and the results are depicted in figure \ref{fig:PEC}\thinspace (b). The lowest PECs for the cation, which are likely to be dominantly populated by the strong-field NIR pulse, have potential wells along this coordinate, these states are predicted to be non-dissociative (figure \ref{fig:PEC}\thinspace (b)). This suggests that the dissociation triggered by the NIR pulse proceeds along a different reaction coordinate(s), probably via more complex structural rearrangements that take significantly longer to evolve.

\section{Conclusions}

The OL-induced dissociation of the prototypical chiral molecule 1-iodo-2-methyl-butane was studied through measuring intensity-dependent fragment yields for femtosecond UV (267~nm) and NIR (800~nm) laser pulses. 
These photoinduced dynamics were also probed via femtosecond time-resolved inner-shell photoionization at the I 4d edge using XUV pulses from FLASH in conjuction with velocity-map imaging. By analyzing the KE distribution of the I$^{2+}$ as a function of different delay times, the dynamics of the neutral and ionic dissociation channels populated by the UV and NIR pulses were explored. Under the employed conditions, the UV pulses predominantly led to resonant single-photon absorption whilst the NIR pulse ionized the molecule, and could initiate dissociative ionization. These dissociative ionization process was found to proceed several times slower than the UV-induced photodissociation. This observation could be rationalized in terms of the weakly bound nature of low-lying PECs of the cation along the C-I stretch coordinate.

\vspace{6pt} 

\section{Author Contributions}
MI conceived and proposed the experiment.
The experiment was performed by VM, FA, PhS, RB, SB, TMB, MBu, SD, PG, DH, DK, ML, JWLL, LM, RM, HO, CP, DRo, KS, LS, IV, RW, VZ, BE, and MI.
The optical laser system was operated and adjusted by BM, RB, BE, and PG with support in the design by AG.
Engineering support including modifications to the spectrometer was provided by BE, and DRa.
Work related to the target was performed by RP, DK, and IV.
VM and FA analyzed the data with further contributions during the beamtime from PhS, LM, CP, and SD.
LI and ZL performed the theoretical calculations.
CvKS has set-up and supported the operation of the polarizing mirrors.
VM, FA LI, PhS, RB, SB, TMB , GB, MBr, MBu, PhD, SD, AE, AG, PG, DH, DK, ML, JWLL, ZL, BM, LM, RM, MM, HO, CP, RP, DRa, DRo, KS, LS, RT, CV, IV, CvKS, RW, PW, VZ, BE, and MI contributed to in-depth discussions and interpretation.
VM, FA, LI, and MI wrote the manuscript with contributions from all authors.

\section{Conflicts of Interest}
The authors declare no competing interest.

\section{Data Availability}
The datasets used and analyzed during the current study are available from the corresponding author upon reasonable request.

\section{Acknowledgments}
We acknowledge DESY (Hamburg, Germany), a member of the Helmholtz Association HGF, for the provision of experimental facilities. Parts of this research were carried out at FLASH and we would like to thank all scientific and technical teams for smooth operation of the machine and assistance in using the BL1 CAMP endstation.  MI, VM, and PhS acknowledge funding by the Volkswagen Foundation for a Peter-Paul-Ewald Fellowship (Grant Number 93285). This work was partly funded by the Deutsche Forschungsgemeinschaft (DFG)–Project No. 328961117–SFB 1319 ELCH (Extreme light for sensing and driving molecular chirality). It has furthermore been supported by the Bundesministerium f\"ur Bildung und Forschung (BMBF) under grant 13K22CHA. VM and MI acknowledge furthermore the support by the Cluster of Excellence 'CUI: Advanced Imaging of Matter' of the Deutsche Forschungsgemeinschaft (DFG) $–$ EXC 2056 – project ID 390715994. LI acknowledges support from DESY (Hamburg, Germany), a member of the Helmholtz Association HGF. SLAC National Accelerator Laboratory is supported by the U.S. Department of Energy, Office of
Science, Office of Basic Energy Sciences under Contract No. DE-AC02-76SF00515. FA, MB, DH, RM and CV acknowledge funding from EPSRC Programme Grants EP/L005913/1, EP/V026690/1 and EP/T021675/1. MBu acknowledges the support from the UK EPSRC (EP/S028617/1 and EP/L005913/1). We furthermore acknowledge the Max Planck Society for funding the development and the initial operation of the CAMP end-station within the Max Planck Advanced Study Group at CFEL and for providing this equipment for CAMP@FLASH. The installation of CAMP@FLASH was partially funded by the BMBF grants 05K10KT2, 05K13KT2, 05K16KT3 and 05K10KTB from FSP-302. SD, KS, LS, and SB acknowledge funding from the Helmholtz Initiative and Networking Fund through the Young Investigators Group Program (VH-NG-1104). ZL was supported by National Natural Science Foundation of China (Grants No. 12174009, 12234002, 92250303). Besides, KS and SB were supported by the Deutsche Forschungsgemeinschaft, project B03 in the SFB 755 - Nanoscale Photonic Imaging. DRo was supported by the US National Science Foundation through grant PHYS-1753324. ML and VZ acknowledge the support from the Swedish Research Council (VR) through the R\"ontgen-\AA ngstr\o m Cluster program (grant No. 2021-05967).

\bibliographystyle{ieeetr}

\end{document}